# Joule heating of dilute 2D holes in a GaAs quantum well

Xuan P. A. Gao[1,2], Allen P. Mills, Jr.[2], Arthur P. Ramirez[3], Steven H. Simon[4], Loren N. Pfeiffer[4] and Kenneth W. West[4]

[1]*Dept. of Applied Physics & Applied Math, Columbia University, New York City, NY 10027*
[2]*Physics Dept., University of California, Riverside, CA 92521*
[3]*Los Alamos National Laboratory, Los Alamos, NM 87545*
[4]*Bell Laboratories, Lucent Technologies, Murray Hill, NJ 07974*

We present measurements of the Joule heating of a 2D hole gas (2DHG) formed in a 30nm GaAs quantum well. The hole density is in the range $(4.6\text{-}18.9)\times 10^9 \text{cm}^{-2}$ and exhibits an apparent metal-to-insulator transition (MIT) with a critical density $6\times 10^9 \text{ cm}^{-2}$. In the limit of zero heating power density $P$, the GaAs lattice is within 2 mK of the 6 mK base temperature of our dilution refrigerator determined by $^3$He melting curve thermometry. Throughout the range of heating power densities used (1 to $10^6$ fW/cm$^2$), the temperature rise of the lattice is estimated to be negligible compared to the temperature rise of the hole gas. We argue that the hole scattering rate is only a function of the hole temperature, with little dependence on the lattice or impurity temperatures in the relevant temperature range below 150 mK. We have therefore made measurements of the hole resistivity at negligible heating power density ($P$<5fW/cm$^2$) as a function of measured lattice temperature in the range 6 to 150 mK. We then use the hole resistivity measured in a cold lattice to estimate the temperature of the 2D hole gas as a function of $P$. In the low hole density insulating phase, the heating power density P(T) that heats the hole gas to a temperature T exhibits a dependence $P \sim T^{\,2}$ for $T$<30mK, gradually changing to $P \sim T^{\,4}$ for higher temperatures. On the metallic side of the MIT, $P$ is proportional to $T^{\,5}$ or $T^{\,6}$ with a magnitude roughly 4 times less than the power density required to heat the insulating phase to the same temperature. Our measurements are within a factor of two of the available quantitative theoretical predictions for the hole energy loss rate as a function of temperature.

**I. Introduction.**

One parameter scaling theory predicts that all two-dimensional systems of non-interacting fermions in zero magnetic field and with finite disorder become insulating as they approach zero temperature [1,2]. Experiments on clean dilute 2D systems, on the other hand, have shown a range of 2D carrier densities for which the resistivity *R(T)* decreases with decreasing temperature *T* [3-6]. Thus the question of the true ground state of strongly interacting 2D systems at zero magnetic field has been actively revisited during the last decade. Although there has been a tremendous amount of experimental and theoretical effort on this topic, there is still no consensus. In some models based on single particle physics, the metallic-like temperature dependent resistivity is attributed to the temperature dependent disorder or temperature dependent screening [7,8]. In these models, the metallic resistivity *R(T)* is predicted to increase again at low enough temperature when the effects of quantum interference become dominant. Testing this prediction involves the measurement of *R(T)* at low *T* where the energy loss rate of 2D carriers is very small. Therefore the thermalization rate of 2D carriers and questions about how far one can cool dilute metallic electrons or holes have become subjects of concern [9].

The 2D electron/hole gas relaxes its energy by emitting phonons. For the electron gas in high mobility GaAs heterostructures at low temperatures in the Bloch-Gruneisen (BG) regime, where the electron-phonon coupling is controlled by screened piezoelectric coupling, it is observed that the acoustic phonon radiation power is proportional to $T^5$ [10,11]. When the disorder becomes strong, the phonon emission rate is enhanced due to the reduced screening seen by the carriers. The phonon emission rate in the low mobility limit is found to be proportional to $T^4$ [12-14]. The change of power law is attributed to the fact that the electrons in a disordered medium cannot move fast enough to respond adiabatically to the electric field generated by phonons and so one has to take into account dynamic screening in the dirty limit instead of static screening, as explained by Chow et al. [12,13]. So far as we know, no phonon emission rate measurements have yet been reported for the 2D hole gas system.

In this paper we present our heating measurements on a dilute 2D hole gas (2DHG) in a 30nm wide quantum well at 6mK bath temperature. The experiment is based on the fact that hole-hole scattering rate is much greater than the inelastic hole-phonon scattering rate for the 2DHG at low temperatures. Thus the hole gas maintains its internal thermal equilibrium as it drifts with weak coupling to the lattice in response to an applied electric field. We thus assume that the heating power density $P$ in equilibrium is proportional to the phonon emission rate $\Gamma$, $\Gamma(T) \propto P$. We find that the phonon emission rate changes from $\Gamma \sim T^5$ or $T^6$ into $\Gamma \sim T^4$ as the 2DHG is tuned from the metallic (high mobility) to the insulating (low mobility) side of the Metal-Insulator transition (MIT). This is consistent with all previous reports on the phonon emission rates for a 2D electron gas. The magnitude of the phonon emission rate of our system in the metallic side of MIT is in good agreement with theoretical expectations. The available theory for phonon emission in dirty limit, however, overestimates the phonon emission rate by at least a factor of two for our system in the insulating state.

## II. Apparatus and Methods.

**Sample.** Our measurements were performed on a back-gated hole-doped GaAs sample made from one of the wafers used in our previous study [5], a (311)A GaAs wafer using $Al_xGa_{1-x}As$ barriers (typical x=0.10) and symmetrically placed Si delta-doping layers above and below a pure GaAs quantum well of width 30 nm. The sample was thinned to ≈150 μm and prepared in the form of a Hall bar, of approximate dimensions (2.5×9) mm$^2$, with diffused In(5%Zn) contacts. The hole-density was varied from 4.6 to $18.9 \times 10^9$ cm$^{-2}$ by means of a gate at the back of the sample. The zero gate bias density was $1.2 \times 10^{10}$ cm$^{-2}$ and the density at which we observe the zero magnetic field (H = 0) metal-to-insulator (MI) transition is roughly $6 \times 10^9$ cm$^{-2}$. Thus we can achieve both the clean limit $ql > 1$ and the strongly disordered limit $ql \ll 1$ for piezoelectric hole-phonon scattering on the same sample, where $q$ is the average magnitude of the phonon wave vector and $l$ is the impurity scattering mean free path of holes. The sample exhibits an apparent metal-to-insulator transition (MIT) at a hole density $p_c = 6 \times 10^9$ cm$^{-2}$, corresponding to a Wigner-Seitz radius $r_s \approx 20$ at $p_c$ assuming an hole effective mass $m^* = 0.18 m_e$ [15]. The measurement current (~100 pA, 2 Hz) was applied along the [2̄33] direction. Independent measurements of the longitudinal resistance per square, $R_{xx}$, from contacts on both sides of the sample were made simultaneously as the temperature was varied. The sample has a low temperature hole mobility $\mu = 2 \times 10^5$ cm$^2$V$^{-1}$s$^{-1}$ for zero gate bias.

The sample was mounted on a copper slug, which was screwed on the copper tail of the mixing chamber of an Oxford TLE-200 top-loading dilution refrigerator with 6mK base temperature. The sample lattice temperature $T_L$ was measured by a germanium resistance thermometer calibrated from the 6 mK base temperature to 320 mK by in situ $^3$He melting curve thermometry, the primary standard at such temperatures [16]. The Ge thermometer was mounted on the mixing chamber, and is thus measuring the refrigerator temperature $T_R$. With commercial electronics, we were unable to obtain measurements on the germanium resistance and the sample resistance below 50mK. We used instead, spatially compact, well-shielded low power custom instrumentation. With this instrumentation we observed that at low carrier densities, for which the sample exhibits insulating behavior, the sample resistivity $R_{xx}(T)$ keeps increasing as $T_R$ is decreased to the base temperature. From this we infer that $T_L$ is not significantly different from $T_R$ at sufficiently low measurement power levels.

**Resistance measurements**.  We measure the sample and Ge thermometer resistances using a set of lock-in type amplifiers assembled into a solid Al box. The measurement leads within the dewar were part copper and part superconductor with a measured resistance 1.4-6.1 Ω per lead and average value $R_{lead}$=3.2 Ω per lead, inductance $L_{lead}$=2.2 µH per lead, and capacitance to ground $C_{lead}$=340 pF. The measurement leads were connected via a Fisher model 105, 18-pin shielded connector on the top of the refrigerator to a 1.6 m length of Belden M-8774 cable made of 9 isolated individually foil-wrapped twisted pair insulated #22 AWG Cu conductors within an outer braided shield grounded to the Fisher connector outer case on one end. The cable was terminated within one section of a solid Al box by hard connections of the outer shield and the ten shields to the box, and of the individual wires to LC feed-through filters ($C_{filter}$=70nF, $L_{filter}$=0.7µH) leading to section two of the box. Section two of the Al box contains a gain of 100 AD795 voltage-follower preamplifier for each of the four voltage measurement leads from the sample and one preamplifier to measure the voltage being applied to the sample. The positive input leads of the first four preamplifiers were attached to nothing else than to the LC feed-through filters through 1kΩ resistors. A square wave modulation voltage was applied to the sample through a resistor $R_V$=10kΩ connected to one of the LC feed-through filters and thence to the 8774 cable. The sample current was detected directly on the negative input lead of a voltage-to-current converter AD795 operational amplifier having a feedback impedance comprised of a 1MΩ resistor in parallel with a 100pF capacitor. We used a 2 Hz square-wave modulation and measurement half cycles consisting of a one-quarter half-period delay followed by a three-quarter half period signal integration time. We used ×10 differential amplifiers [AD706] to subtract pairs of preamplified signal voltages. We used MOSFET switches [DG442] to square-wave modulate the excitation voltage derived from a floating 3.6V Li battery and to demodulate the ×1000 preamplified signal using a single integrator to average the difference between the positive and negative signal amplitudes during the later three fourths of each half measurement cycle. The overall gain of each channel is $8.15\times10^3$ and the offset errors are less than 1 mV, thus allowing us to make useful measurements using modulation currents as low as 5 pA. Section three of the Al box is separated from section two by LC feed-through filters (1µF/7µH) and contains a 4 MHz clock and TTL drivers for the modulators as well as (1 MΩ in series followed by 2 µF to ground) output filters connected to the BNC output connectors. The output voltages are measured by HP 34401A digital voltmeters connected via a GPIB bus to a personal computer using LabView. The sample resistances are computed from the ratio of the voltage and current output signals and the known sample geometry.

The resistance of the Ge thermometer is determined by a four-wire measurement, so that the resistance of the measurement leads does not affect the measurement. However, unlike the GaAs quantum well samples that are the subject of our experiment, the Ge thermometer has only two connections. This of course means that the measured resistance is affected by the Ge contact resistance, but this is of no consequence because the latter is present during the calibration of $R_{Ge}(T)$ also. Having only two contacts is actually very important for our ability to measure the Ge resistance at low temperatures where $R_{Ge}$ can be greater than 100 MΩ. For such a resistance, measurement by a four wire method with four separate contacts might be difficult to perform at a few Hz modulation rate given the 1µF filter capacitors we are using. However, having only two contacts means that the Ge resistor can be connected to virtual ground at all contacts so that there are no long time constants. The Ge resistor is excited through one contact from a 10 kΩ, 2 Hz voltage source. The other contact to the Ge resistor is connected to a virtual ground, the negative input lead of a voltage to current converter AD797 operational amplifier having a 50 MΩ in parallel with 10 pF feedback impedance.

**External noise sources**. The noise power delivered to the sample due to its being connected to the measuring apparatus is calculated as follows. The Johnson noise power from each of the six measurement leads attached to the sample is $p_J=4kT_{lead}fR_{lead}/R_{samp}=[R_{lead}/R_{xx}]\times 6.4\times 10^{-14}$ W, where $T_{lead}\approx 200$K, f=8MHz and where the sample resistance is about twice the sample resistivity $R_{xx}$ given in ohms per square.

Normalizing to the 0.23 cm$^2$ total sample area, the total Johnson noise input to the sample from the six leads is $P_J=[3k\Omega/R_{xx}]\times 1.1\times 10^{-15}$ W/cm$^2$. This is the only significant source of Johnson noise and raises the temperature of the holes in the quantum well of our sample by less than 2 mK for all values of $R_\square$ encountered in this study. The $R_V=10k\Omega$ resistor in series with the applied modulation voltage applies a Johnson noise power to the sample of $p_V=kT/[\pi C_{filt}R_V]\times[R_V/(R_V+R_{samp})]=R_V/(R_V+R_{samp})\times 1.6\times 10^{-18}$ W, so that the noise power per unit area is $P_V<7\times 10^{-18}$ W/cm$^2$. The current measuring operational amplifier has an input voltage noise of 5 nV per root hertz at frequencies above 100 Hz. The relevant bandwidth is $f=1/2\pi C_{filt}R_{samp}$ so the power delivered to the sample is $p_I=(5\times 10^{-9}\text{V per root Hz})^2/(2\pi C_{filt}R_{samp}^2)=[5k\Omega/R_{xx}]\times 0.6\times 10^{-18}$ W, or $P_I=[3k\Omega/R_{xx}]\times 5\times 10^{-18}$ W/cm$^2$.

**Carrier density**. We determined our carrier density from Shubnikov-de Haas (SdH) magnetoresistance measurements obtained by averaging the longitudinal resistances per square $R_{xx1}$ and $R_{xx2}$ obtained from both sides of the sample. We plot the SdH minima vs. gate bias and fit the data using a straight line $p(V_{gate})=p_0-\alpha V_{gate}$ and the relation $p(V_{gate})=(2.418\times 10^{10}$ cm$^{-2})\times H_{\nu=1}(V_{gate})$. The measured slope is $\alpha=(0.420\pm 0.005)\times 10^9$ cm$^{-2}$V$^{-1}$ independent of sample history and the gate capacitance is $e\alpha=(67.3\pm 0.7)$pFcm$^{-2}$. The zero gate bias density is $p_0=(12.0\pm 0.5)\times 10^9$ cm$^{-2}$, where the variation reflects a gradual increase in the zero gate bias density over the course of six months time.

**Temperature calibration.** The Ge resistance thermometer was calibrated by Lake Shore Cryotronics, Inc. above 50mK. Below 50mK we used a $^3$He melting curve thermometer (MCT) to calibrate the Ge resistor. Our MCT is similar with the type that Greywall *et al.* used in Ref.16. A $^3$He sample is confined in the $^3$He chamber of the MCT and kept at a constant average density by a capillary blocked by frozen solid $^3$He. The pressure of the $^3$He solid and liquid mixture in the chamber is sensed by a parallel plate capacitor through a diaphragm on the $^3$He chamber. The pressure $P_{MCT}$ vs. capacitance $C_{MCT}$ calibration of the MCT capacitance manometer was obtained at 1.5K by reading a Heise bourdon tube pressure gauge and a capacitance bridge connected to the MCT after cycling the MCT three times from 0 to 35 Bar to reduce hysteresis. We calibrated the Heise bourdon gauge against a 0-35 Bar full scale MKS Instruments capacitance manometer, which has a stated 0.3% of full scale limit of error and 3mBar readability (MKS model 722ARCTCD2FK). The calibration data of the Heise gauge pressure $P_{Heise}$ vs. the MKS manometer pressure $P_{MKS}$ are plotted in Fig.1a with a linear fit $P_{Heise} = \Delta P+ B\times P_{MKS}$ where $\Delta P=35.36\pm 3.1$mBar and $B=0.99681\pm 0.00017$. The residual errors of the MKS manometer-Heise gauge calibration are plotted as Fig.1b. From the fitting result we see that our Heise gauge has an accuracy comparable to the MKS capacitance manometer. Since the MKS-Heise gauge calibration is within the error limit of the MKS manometer, we simply took the MCT pressure $P_{MCT} = P_{Heise}- \Delta P$.

The calibration data of $P_{MCT}$ vs. the MCT capacitance $C_{MCT}$ are shown in Fig.1c, which includes all the data from seven sets of calibration during a two month time period. The function

$$P_{MCT}(C_{MCT}) = P_0+ k_1 C_{MCT}^{-1} +k_2 C_{MCT}^{-2} \qquad (1)$$

was used to fit the $P_{MCT}(C_{MCT})$ calibration data of the MCT, yielding $P_0 = 47.09032\pm 0.01402$, $k_1 = -200.84704 \pm 0.19805$ and $k_2 = -22.4238 \pm 0.65098$. Fig.1d plots the corresponding residual errors. The MCT was cooled with initial $^3$He pressure above 34 Bar to ensure there would be $^3$He solid in the MCT cell at the lowest temperature. During the cooling procedure, the MCT capacitance $C_{MCT}$ and Ge resistance $R_{Ge}$ were recorded. From $C_{MCT}$, the corresponding $^3$He pressure in MCT was calculated by Eq.1. The value $P_{MCT}$ was then converted into a standard temperature from the known Greywall $^3$He melting pressure-temperature scale in Ref.16. Fig.1e shows the $C_{MCT}$ versus absolute temperature. The pressure (capacitance) minimum value is $29.3465\pm 0.0005$Bar. This pressure minimum is $30.5\pm 0.5$mBar higher than the value of Ref.16. This is consistent with the fitted zero offset of the Heise vs. MKS gauge calibration, $\Delta P=35.36\pm 3.1$mBar. The pressure is minimal at 312mK(Lakeshore reading), 2% lower than the 318mK given by Ref.16. The origin of this 2% deviation is not clear, but does not affect the calibration much since it is the accuracy of the $^3$He pressure that determines the error of the calibrated temperature. We note that the 6mK temperature difference at the pressure minimum here and Ref.16 was not caused by the MCT not being in thermal equilibrium with the refrigerator because the pressure minimum was passed through several times. The $^3$He pressure at base temperature was 34.220Bar, corresponding to a 6mK base temperature. The standard deviation of our $P_{MCT}(C_{MCT})$ calibration, $\varepsilon = \pm$ 14mBar. The error of $P_{MCT}$ at 34.220Bar of the Heise gauge calibrated at 29.316Bar is estimated to be less than $(34.220-29.316) \times 0.3\%$ Bar $\approx$ 15mBar. Thus we estimate the error of $P_{MCT}$ at base temperature, $\delta P_{MCT} < \pm (15+\varepsilon)$ mBar $= \pm$ 29mBar, where we are calculating a conservative estimate for $\delta P_{MCT}$ by adding the error contributions rather than their squares. This corresponds to a base temperature error $\delta T = \delta P_{MCT} \div dP_{MCT}/dT = 0.029/40.01 = 0.7$mK, with $dP_{MCT}/dT = -40.01$BarK$^{-1}$ at 6mK given by Ref.16. We present the calibrated Ge thermometer resistance from 100mK to 6mK in Fig. 2. Note that the Ge resistance we calibrated by $^3$He melting curve thermometry is well reproducible and consistent with the $T$ >50mK calibration provided by Lake Shore Cryotronics, Inc.

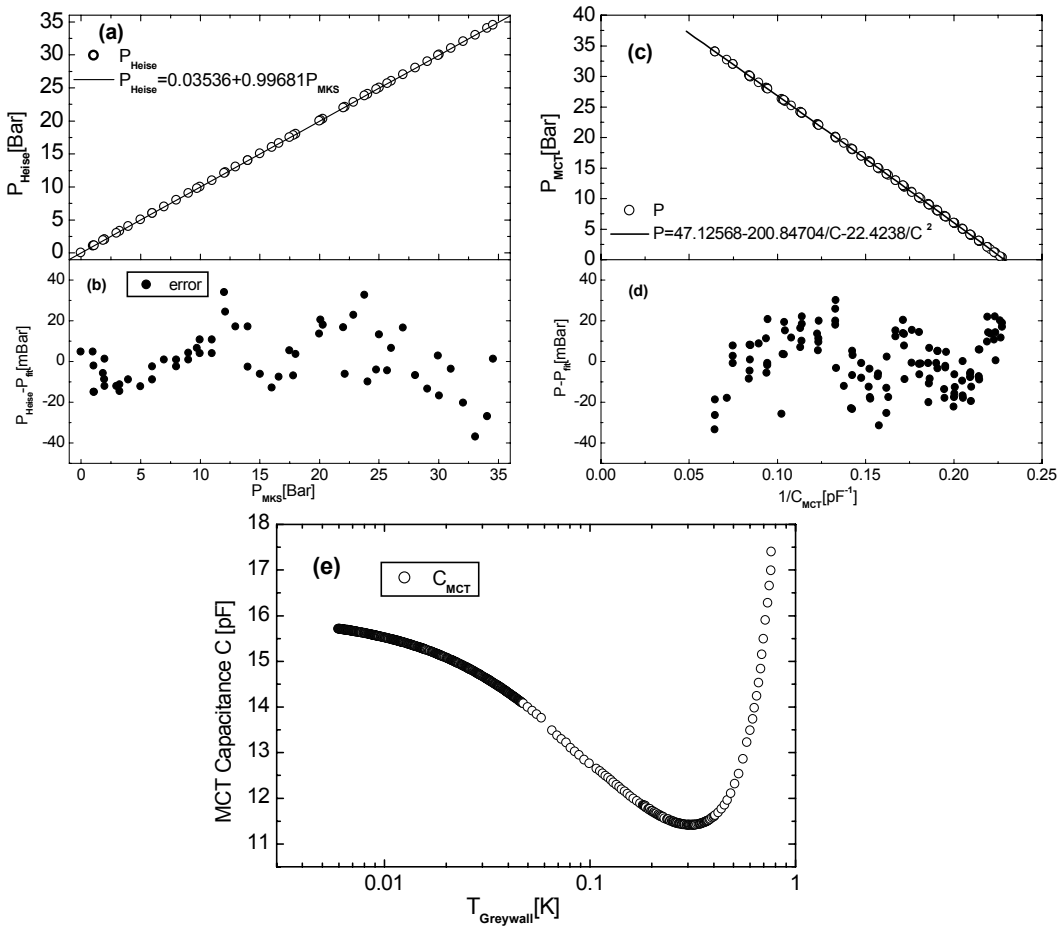

*Figure 1.* (a) The calibration of the Heise gauge pressure $P_{Heise}$ vs. $P_{MKS}$, the MKS capacitance manometer pressure. The MKS capacitance manometer has ±0.3% accuracy. The calibration data are fitted with linear dependence. (b) The difference between the data and the linear fit in (a). (c) Pressure vs. inverse capacitance calibration data of the $^3$He melting curve thermometer (MCT), fitted by $P_{MCT}(C_{MCT}) = P_0 + k_1 C_{MCT}^{-1} + k_2 C_{MCT}^{-2}$. (d) Residual error between the fit and the data of (c). (e) Melting curve thermometer capacitance versus absolute temperature

calibrated via Greywall's method. That 9 data points between 60 and 100 mK appear to be deviating from the expected line is attributed to the fact that the refrigerator is cooling very rapidly in this temperature region so that thermal equilibrium has not been attained. At temperatures below 50mK there was adequate time to establish thermal equilibrium between the refrigerator, Ge thermometer and the MCT.

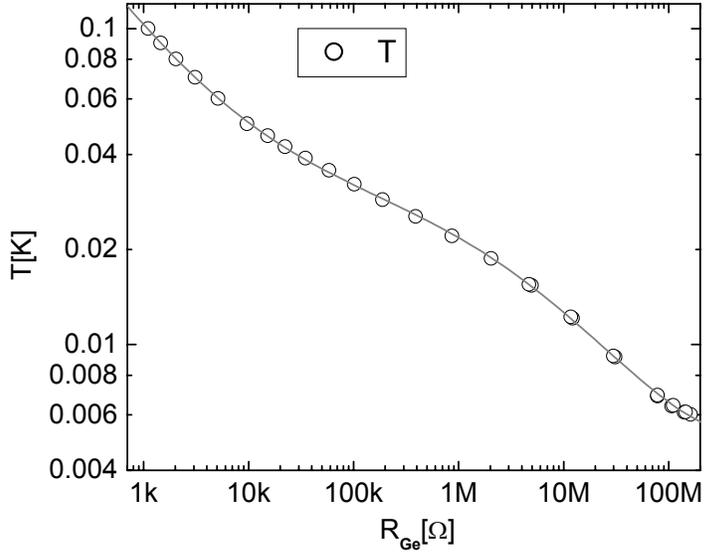

*Figure 2*. Ge thermometer resistance versus $^3$He melting curve thermometry temperature. The gray line connecting the data is a fifth order polynomial fit, $y = a_0+a_1x+a_2x^2+a_3x^3+a_4x^4+a_5x^5$ with $y=\log T$, $x=\log R_{Ge}$, $a_0$= -1.55703, $a_1$= 1.91515, $a_2$= -1.16154, $a_3$= 0.27138, $a_4$= -0.02865 and $a_5$= 0.00112.

### III. Measurements.

Fig. 3a shows a measurement of longitudinal resistance per square $R_{xx}(T)$ from 6 to 200 mK for various gate biases corresponding to a carrier density range $(4.6-18.9)\times10^9$cm$^{-2}$. The data were obtained with a 25 Gauss magnetic field applied to cancel the residual field of the superconducting solenoid. The excitation levels were limited below 5 fW/cm$^2$ in all cases to reduce heating of the holes. Fig. 3b presents the measurement of $R_{xx}$ versus the driving signal power $P$ with the sample sitting at 6mK base temperature.

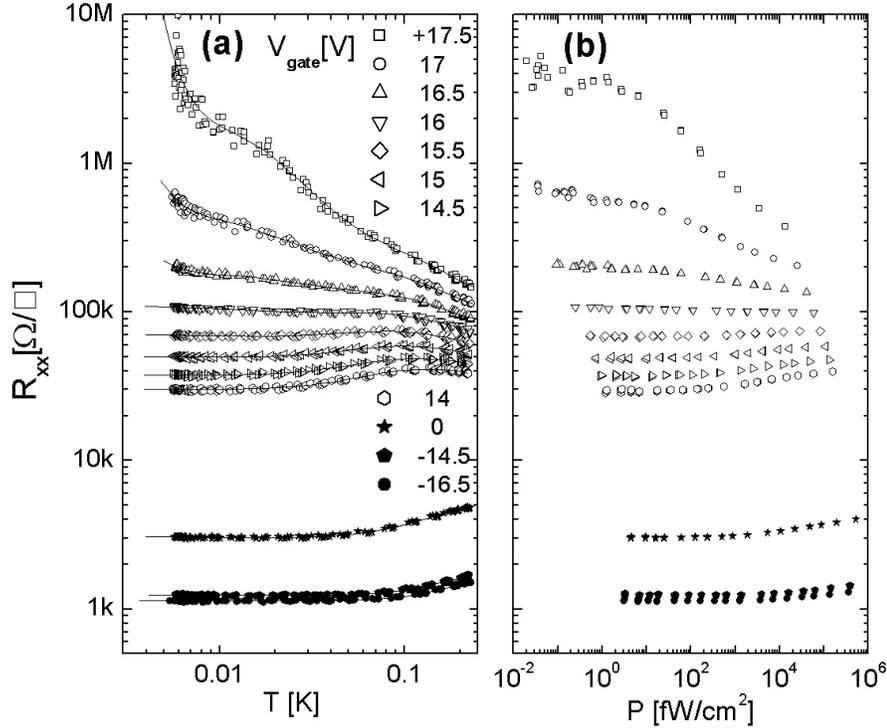

*Figure 3.* a) Longitudinal resistance versus temperature for 2D hole gas in a 30nm GaAs quantum well at various gate biases Vgate. The hole density p at Vgate(Volts) is given by p=12.0-0.42Vgate [$\times 10^9 cm^{-2}$]. The measurement power is less than 5fW/cm$^2$. The solid curves are 6$^{th}$ order polynomial fits. b) Longitudinal resistance versus excitation power per unit area with the sample held at the base temperature 6 mK.

In order to determine the energy relaxation rate of the hole gas, we studied the apparent nonlinear resistance of our sample at zero magnetic field. Interpreting the excitation power dependent nonlinear resistance as being caused by $I^2R$ heating of the holes, we convert the $R_{xx}(P)$ and $R_{xx}(T_{Ge})$ data of Figs 3a and 3b to $P(T_{eff})$ exhibited in Fig 4. The solid lines in Fig 3a are 6th order polynomial approximations to u(v) where u=lnR and v=lnT. To produce the points $P(T_{eff})$ in Fig 4, the effective temperature $T_{eff}$ corresponding to each given power $P$ in the $R_{xx}(P)$ measurements of Fig 3b is determined by looking up the value of $R_{xx}$ in the polynomial curves representing $R_{xx}(T)$. Due to the flatness of the $R_{xx}(T)$ curves at low temperatures we are not able to get $P(T_{eff})$ below 20mK for $V_{gate}$=0,-14.5 and –16.5 volts. In the clean limit $ql>1$, $P(T) \sim T^5$ is expected for the screened piezoelectric coupling mediated acoustic phonon emission from 2D electron/hole gas. From the estimated thermal phonon wave vector $q=k_BT/s\hbar$ and hole mean free path $l = v_F\tau =v_Fm^*/(pe^2R_{xx})$ we find $ql>1$ is always satisfied for the three curves with highest carrier densities in Fig 4 [17]. Indeed, for the three densities deep in the metallic side of MIT in Fig. 4 P(T) may be interpreted as crossover from a $T^5$ dependence to $T^6$ dependence in the temperature range we studied. In the critical region or the insulating side of the MIT, it can be seen from Fig. 4 that P(T) changes into power law lower than $T^5$. The temperature dependence of the cooling rate of the holes given a lattice temperature $T_L$ for $V_{gate} \geq 14$ volt can be described by the form [12,13]

$$P(T)=b_2^2(T^2-T_L^2)+ b_4^4(T^4-T_L^4) \tag{2}$$

where we include a $T^2$ term in Eq. (2) to account for the cooling due to the leads attached to the sample [18].

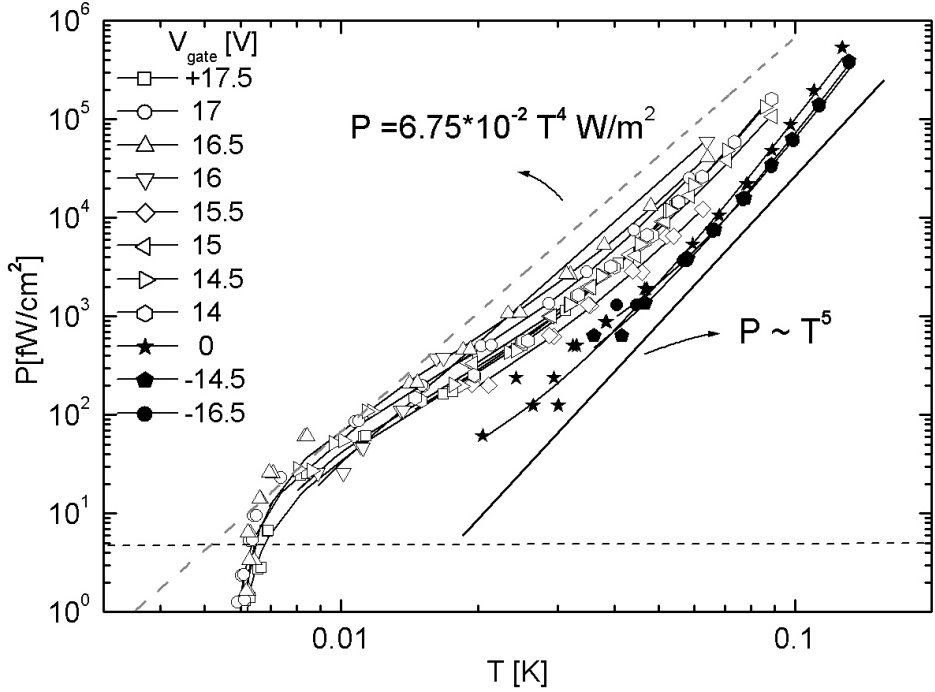

*Figure 4.* Excitation power per unit area, P, corresponding to hole temperature T inferred from Fig 3 (a) and (b) for a lattice temperature of 6 mK,. The dashed horizontal line represents the maximum power level used in obtaining the data of Fig. 3. The dashed gray line is the expected P(T) by Eq. 4 for $R_{xx}=h/e^2$. $P \sim T^5$ is also drawn as a black line for reference.

**IV. Discussion and Conclusion.**

We now turn to the quantitative discussion of the data. The theoretical calculation of the acoustic phonon emission power in a clean 2D electron gas with density n is [Eq. 1 of Ref. 13]

$$P = [1.65 \text{ fW/cm}^2] \times [T/10 \text{ mK}]^5 \times [n/10^{10} \text{ cm}^{-2}]^{-1/2}. \tag{3}$$

There have been several experiments on the 2D electron gas in GaAs supporting Eq. 3 (e.g. Ref.11 and Ref.14). Assuming Eq. 3 also describes holes, we would expect $P=1.51\times10^{-1} T^5$ (W/m$^2$) for $V_{gate}=0$ of our sample. Fitting the data in Fig.4 with $P=bT^5$ we get $P=(1.47\pm0.05)\times10^{-1} T^5$ (W/m$^2$) for $V_{gate}=0$, which is in good agreement with Eq. 3. Fitting the $V_{gate}=$ -14.5 and -16.5 data also gives *b* within 30% of the predictions of Eq. 3.

For the *ql <1* regime, the dynamically screened piezoelectric coupling governed phonon emission rate is expected to be [12]

$$P = [67.5 \text{ fW/cm}^2] \times [R_{xx} \times (e^2/h)] \times [T/10 \text{ mK}]^4 \tag{4}$$

Our analysis shows that Eq. 4 overestimates the energy loss rate by about a factor of two for our sample as illustrated by the comparison between the data and the gray line in Fig. 4 representing the prediction of Eq. 4 for $R_{xx} = h/e^2 = 25813 \text{ }\Omega/\square$.

**Minimum hole temperature and weak localization.** We now briefly remark on how cold the holes are and discuss possible weak localization corrections to $R_{xx}(T)$. The continuous changing of $R_{xx}(T)$ in the high resistance region shows that the holes are still cooling at the 6mK base temperature in the insulating state. On the metallic side of the MIT, $R_{xx}(T)$ universally flattens at low temperatures. Simply extrapolating the $T^5$ dependent phonon emission power for low resistance $R_{xx}(T)$ we may infer that the holes are cooled to 10mK with driving powers less than 5fW/cm$^2$ when the sample temperature is 6mK. The extra cooling $T^2$ power from the leads further reduces the minimum estimated hole temperature by a few mK in the low resistance region. Taking into account the $T^2$ electronic cooling from leads, we estimate that the minimum temperature of the holes is below 7.5mK for g<10, where g is the dimensionless conductivity in units of $e^2/h$. Our recent observation of a continuously increasing $R_{xx}(T)$ in the k$\Omega/\square$ regime in a parallel magnetic field supports our contention that the holes are actually cooling below 10mK when the driving power is less than a few fW/cm$^2$ [19]. Weak localization (WL) theory predicts a universal logarithmic temperature dependent conductivity correction to $R_{xx}(T)$ for 2D transport in the weak disorder limit ($k_F l > 1$) at low temperatures. In the resistivity measurement, WL is reflected as a logarithmically increasing resistivity proportional to the resistivity at which WL begins [2]. Our sample exhibits a flattening $R_{xx}(T)$ below temperatures as low as 0.5% of the Fermi temperature of the 2DHG for densities p>p$_c$ with no indication of turning up again [20]. This is in contrast to the prediction of conventional WL theory for Fermi liquids. However, it remains curious that in some studies an upturn in $R_{xx}(T)$ at the lowest temperatures was observed together with a metallic-like $R_{xx}(T)$ at intermediate temperature range for holes in GaAs heterostructures with densities in order of $10^{11}$ cm$^{-2}$ or less [21, 22]. The negative magnetoresistance in the 2D metallic phase was also taken as an evidence for a persistent weak localization effect for a 2DHG in GaAs with high $r_s$ [21, 22].

In conclusion, we have measured the energy relaxation rate in a dilute 2DHG system exhibiting an apparent MIT. The phonon emission power changes from $P \sim T^5$ or $T^6$ to $P \sim T^4$ when the system crosses from the metallic into the insulating state. This is qualitatively consistent with previously reported phonon emission measurements of 2D electron gas in GaAs, where the disorder-induced change in the screening affects the phonon emission character. In the clean limit ($ql > 1$), the phonon emission rates we measured are in good agreement with theoretical values. In the $ql < 1$ regime the phonon emission rate of our system is at least a factor of two smaller than the predictions of available models.